\documentclass[useAMS,usenatbib]{mn2e}
\usepackage{graphicx}
\usepackage{amsmath,amssymb}
\usepackage{bm}
\usepackage{color}
\usepackage{float} 

\def\be{\begin{equation}}
\def\ee{\end{equation}}
\def\ba{\begin{eqnarray}}
\def\ea{\end{eqnarray}}

\begin{document}
\label{firstpage}

\title{Searching for primordial magnetism with multi-frequency CMB experiments}
\author[Levon Pogosian]{%
Levon Pogosian%
\thanks{E-mail: \texttt{levon@sfu.ca}}\\%
Department of Physics, Simon Fraser University, Burnaby, BC, V5A 1S6, Canada}


\maketitle

\begin{abstract}
Bounds on the amplitude of a scale-invariant stochastic primordial magnetic field (PMF) can be significantly improved by measurements of the Faraday Rotation (FR) of CMB polarization. The mode-coupling correlations induced by FR make it possible to extract it from cross-correlations of the B-mode polarization with the E-mode and the temperature anisotropy. In this paper, we construct an estimator of the rotation measure that appropriately combines measurements of the FR from multiple frequency channels. We study the dependence of the signal-to-noise in the PMF detection on the resolution and the noise of the detectors, as well as the removal of the weak lensing contribution and the galactic FR. We show that a recently proposed space-based experiment PRISM can detect magnetic fields of 0.1 nano-Gauss strength at a 2 sigma level. Higher detection levels can be achieved by reducing the detector noise and improving the resolution or increasing the number of channels in the 30-70 GHz frequency range.
\end{abstract}

\begin{keywords}
cosmology -- primordial magnetic fields, CMB polarization, Faraday Rotation
\end{keywords}

\section{Introduction}

Magnetic fields exist in all gravitationally bound structures in space, from planets and stars to galaxies and clusters. While in many cases they are generated by plasma dynamics, explaining the micro-Gauss strength fields observed in galaxies is challenging without a seed primordial magnetic field (PMF) \citep{2002RvMP...74..775W,2013A&ARv..21...62D}. Giving additional impetus to the PMF hypothesis is the claimed detection of a magnetic field in the intergalactic space \citep{2010Sci...328...73N} based on non-observation of GeV $\gamma$-rays expected to accompany TeV blazars.  Further, \citet{2013arXiv1310.4826T} recently reported hints of left-handed magnetic helicity on $10$ Mpc scales measured using $\gamma$-ray data from the Large Area Telescope onboard the Fermi satellite \citep{2009ApJ...697.1071A}. Detecting a PMF would bring us closer to a complete theory of the early universe \citep{2001PhR...348..163G}. Candidate mechanisms for generation of a PMF include inflationary scenarios \citep{1988PhRvD..37.2743T,1992ApJ...391L...1R} and phase transitions \citep{1991PhLB..265..258V}. 

A PMF would imprint signatures in the cosmic microwave background (CMB) carrying valuable clues about its origin. Fields produced in phase transitions have most of the power concentrated near a small cutoff scale set by the plasma conductivity \citep{2003JCAP...11..010D,2011PhRvD..83j3005J}, with practically no strength on cosmological scales that would translate into observable correlations in CMB temperature and polarization. Instead, CMB constraints on such fields come from spectral distortions caused by the magnetic energy being damped into plasma on small scales \citep{2000PhRvL..85..700J,2013arXiv1309.7994K} and from the enhanced recombination rates \citep{2011arXiv1108.2517J}. In contrast, CMB temperature and polarization fluctuations provide competitive bounds on a {\it scale-invariant} PMF \citep{2009MNRAS.396..523P, 2013PhLB..726...45P}. Inflation naturally generates a scale-invariant magnetic field, but the conformal invariance of the electromagnetic action implies an almost complete dilution of its amplitude. One way to circumvent this conclusion is to break the conformal invariance, {\it e.~g.} by coupling the electromagnetic field to the inflaton \citep{1988PhRvD..37.2743T,1992ApJ...391L...1R}. Recently, \citet{2013arXiv1309.3435A} (see also \citep{2013PhRvL.111f1301C}) suggested that a sizeable magnetic field can be generated during a phase of slow-roll inflation by the conformal anomaly, without the need for breaking the conformal invariance in the classical action. 

The physical magnetic field strength scales with the expansion factor $a$ as ${\bf B}^{\rm phys}={\bf B}/a^2$, where {\bf B} is the ``comoving'' strength. A common measure of a tangled PMF is its comoving amplitude smoothed over a length $\lambda$, $B_\lambda$. For scale-invariant fields, this quantity is independent of $\lambda$ and is the same as $B_{\rm eff} \equiv \sqrt{8\pi \epsilon_B}$, where $\epsilon_B$ is the total magnetic energy density. Thus, we will use $B_{\rm SI}=B_\lambda=B_{\rm eff}$ to denote the strength of a scale-invariant PMF. 

Recently released Planck data limits $B_{\rm SI}$ to a few nano-Gauss (nG) \citep{2013arXiv1303.5076P}, although these results assume initial conditions in which the magnetic stress-energy is compensated by that of the surrounding fluid. There are additional modes, such as the ``passive mode'' \citep{2004PhRvD..70d3011L,2010PhRvD..81d3517S,2010JCAP...05..022B} generated prior to neutrino decoupling, and the so-called  ``inflationary magnetic mode'' \citep{2012PhRvD..86b3519B,2013PhRvD..88h3515B} produced during the inflation. The amplitudes of these modes depend on the details of the inflationary magnetogenesis and accounting for them can significantly affect the CMB constraints \citep{2013PhRvD..88h3515B}. Similar, {\it i.~e.} a few nG, level constraints on $B_{\rm SI}$ were obtained by \citet{2013ApJ...770...47K} from Lyman-$\alpha$ spectra \citep{2002ApJ...581...20C}. 

The bounds mentioned above are based on the effects of the magnetic filed stress-energy, which is {\it quadratic} in the magnetic field strength. Since the CMB and matter power spectra are, in turn, quadratic in stress-energy, the PMF contribution to them scales as $(B_{\rm SI})^4$. Moreover, the PMF contribution to the CMB temperature and polarization spectra comes as a small addition to a stronger signal. Thus, improving current CMB bounds on $B_{\rm SI}$ by an order of magnitude considering solely the effects of the magnetic stress-energy would require a $10^4$-fold improvement in the accuracy of the measured spectra and the theoretical modelling of all other sources that contribute to them. This will not happen in the foreseeable future. 

On the other hand, Faraday Rotation (FR) of CMB polarization is {\it linear} in $B_{\rm SI}$ and offers a way to detect sub-nG strength PMF with the next generation of experiments~\citep{2009PhRvD..80b3009K,2012PhRvD..86l3009Y,2013PhRvD..88f3527D}. Prior studies~\citep{2012PhRvD..86l3009Y,2013PhRvD..88f3527D} provided estimates at a single frequency and, while improvements due to measurements at more frequencies~\citep{2009PhRvD..80b3009K} were considered, they were not studied with due care until now. In this work, we generalize the estimator of the FR angle at a single frequency used in~\citep{2012PhRvD..86l3009Y,2013PhRvD..88f3527D} to that of a rotation measure that appropriately accounts for the covariance of measurements of the FR at multiple frequencies.

\section{Estimator of rotation measure from multiple maps}

When CMB radiation passes through ionized regions permeated by magnetic fields, its polarization is rotated by an angle \citep{1996ApJ...469....1K,1997PhRvD..55.1841H}
\begin{equation}
\alpha(\hat{\bf n}) = \frac{3c^2 \nu_0^{-2}}{{16 \pi^2 e}}
\int \dot{\tau} \ {\bf B} \cdot d{\bf l} 
=c^2 \nu_0^{-2} \ {\rm RM}(\hat{\bf n}) \ ,
\label{alpha-FR}
\end{equation}
where $\hat{\bf n}$ is the direction of the line of sight, $\dot{\tau}$ is the differential optical depth, $\nu_0$ is the detector frequency, ${\bf B}$ is the comoving magnetic field strength, and $d{\bf l}$ is the comoving length element along the photon trajectory. Measurements of FR at different frequencies probe the same rotation measure (RM), which is the quantity of interest for constraining the PMF. Eq.~(\ref{alpha-FR}) implies that a significant RM can be produced by a small PMF over a large distance, which is the case at recombination, or by a larger magnetic field over a smaller path, which is the case inside our galaxy. The latter acts as noise in the estimates of the RM due to a PMF and was studied in \citet{2013PhRvD..88f3527D}.

The physical interpretation of CMB polarization is aided by separating it into parity-even and parity-odd patterns -- the so-called E- and B-modes~\citep{1997PhRvL..78.2058K,1997PhRvL..78.2054S}. E-modes are expected, since a non-zero intensity quadrupole at recombination necessarily leads to their generation via Thomson scattering, and have been observed \citep{2002Natur.420..772K}. B-modes, on the other hand, can only be produced by sources with parity-odd components, such as gravitational waves \citep{1993PhRvL..71..324C},  topological defects \citep{1997PhRvL..79.1615S} or magnetic fields \citep{2001PhRvL..87j1301S}. FR converts some of the E-mode into B-mode, namely, for small rotation angles, the relation between the spherical harmonic coefficients of E, B and $\alpha$ can be written as \citep{2009PhRvL.102k1302K}
\be
B_{lm}=2\sum_{LM}\sum_{l' m'}\alpha_{LM} E_{l' m'} 
\xi_{lml'm'}^{LM}H_{ll'}^L \ ,
\label{eq:blm}
\ee
where $\xi_{lml'm'}^{LM}$ and $H_{ll'}^L$ are related to Wigner $3$-$j$ symbols:
\ba
\nonumber
\xi_{lml'm'}^{LM} \equiv (-1)^m \sqrt{ (2l+1)(2L+1)(2l'+1) / 4\pi} \\
\times \left(
\begin{array}{ccc}
l  & L  & l'  \\
-m  & M  & m'    
\end{array}
\right) ; \ H_{ll'}^L \equiv 
\left(
\begin{array}{ccc}
l  & L  & l'  \\
2  & 0  & -2    
\end{array}
\right) \ ,
\ea
and the summation is restricted to {\it even} $L+l'+l$. Weak lensing (WL) of CMB by gravitational potentials along the line of sight also converts E- into B-mode \citep{2002ApJ...574..566H}, but couples the odd sums of the modes, making it orthogonal to the FR effect.

Eq.~(\ref{eq:blm}) implies correlations between multipoles of E- and B-modes. Since the CMB temperature (T) and E are correlated, FR also correlates T with B.  The rotation angle can be extracted from EB and TB correlations. Given measurements of B- and E-modes at frequencies $i$ and $j$, respectively, the quantity \citep{2007PhRvD..76j3529P,2009PhRvL.102k1302K,2009PhRvD..79l3009Y,2009PhRvD..80b3510G,2012PhRvD..86j3529G,2012PhRvD..86h3002Y}
\be
[{\hat \alpha}_{B^iE^j,LM}]_{ll'} = {2\pi \sum_{mm'} B^i_{lm}E_{l'm'}^{j*} \xi_{lml'm'}^{LM} \over (2l+1)(2l'+1)C_l^{EE}H_{ll'}^L}
\label{alphallpr}
\ee
provides an unbiased estimator of $\alpha_{LM}$ at frequency $\nu_i$. The corresponding estimator for the RM, which is the achromatic  quantity proportional to the PMF, is
\be
[{\hat r}_{B^iE^j,LM}]_{ll'} = c^{-2} \nu^{2}_i [{\hat \alpha}_{B^iE^j,LM}]_{ll'} \ .
\ee
Note that $[{\hat r}_{B^iE^j,LM}]_{ll'}$ is not symmetrical under interchange of $l$ and $l'$, and one should consider separately contributions from BE and EB correlations. Analogous quantities can also be constructed from products of T and B. Hence, given maps of T, E and B at a number of frequencies (labeled by indices $i,j$), one considers contributions from quadratic combinations 
\be
A \in \{E^iB^j, B^iE^j, T^iB^j, B^iT^j\} \ .
\label{A-quad}
\ee 
The minimum variance estimator ${\hat r}_{LM}$ is obtained by combining estimates from all $A$, accounting for the covariance between them. We derive the variance in ${\hat r}_{LM}$ by closely following the analogous derivation for the rotation angle given in \citet{2009PhRvD..80b3510G}.

The variance in ${\hat r}_{LM}$, for a statistically isotropic RM, is defined as
$\langle {\hat r}^*_{LM} {\hat r}_{L'M'} \rangle =\delta_{LL'} \delta_{MM'} [C_L^{\rm RM}+\sigma^2_{{\rm RM},L}]$,
where $C_L^{\rm RM}$ is the RM power spectrum that receives contributions from the PMF and the galaxy, while $\sigma^2_{{\rm RM},L}$ is the combined variance of individual estimators $[{\hat r}_{B^iE^j,LM}]_{ll'}$. Using notation similar to that of \citet{2009PhRvD..80b3510G}, we can write
\be
\sigma^{-2}_{{\rm RM},L} = \sum_{l, l' \ge l} G_{l l'} 
\sum_{A,A'} [({\cal C}^{l l'})^{-1}]_{AA'} \ 
Z_{l l'}^A Z_{l l'}^{A'} 
\ ,
\label{eq:ebnoise}
\ee
where the sum is restricted to even $l+l'+L$, $G_{l l'} \equiv (2l+1)(2l'+1) (H^L_{l l'})^2 / \pi$, $A$ and $A'$ label the relevant quadratic combinations of E, B and T listed in (\ref{A-quad}), 
\begin{align}
Z_{l l'}^{X^iB^j} = c^2\nu_j^{-2} W^{ij}_{l l'} C^{XE}_l  , \\
Z_{l l'}^{B^iX^j} = c^2\nu_i^{-2} W^{ij}_{l l'} C^{EX}_{l'} ,
\end{align}
with $X$ denoting either $T$ or $E$, and $W^{ij}_{l l'} \equiv \exp[-(l^2+l'^{2}) \theta^2_{ij}/16\ln 2]$ accounts for the finite width of the beam. We take $\theta_{ij} = \max[{\theta^i_{\rm fwhm},\theta^j_{\rm fwhm}}]$, where $\theta^i_{\rm fwhm}$ is the full-width-at-half maximum (FWHM) of the Gaussian beam of the $i$-th channel. The covariance matrix elements, $[{\cal C}^{l l'}]_{AA'}$, are
\begin{align}
[{\cal C}^{l l'}]_{X^iB^j,Y^kB^n}= {\tilde C}^{X^iY^k}_l {\tilde C}^{B^jB^n}_{l'} \\
[{\cal C}^{l l'}]_{B^iX^j,B^kY^n}= {\tilde C}^{B^iB^k}_l {\tilde C}^{X^jY^n}_{l'}
\end{align}
with $X$ and $Y$ standing for either E or T, and
\be
{\tilde C}^{X^iY^j}_l \equiv C^{XY, {\rm prim}}_l+f_{\rm DL}C^{XY, {\rm WL}}_l+ \delta_{X^iY^j} \sigma^2_{P,i} \ ,
\label{clvariance}
\ee
is the measured spectrum, that includes the primordial contribution $C^{XY, {\rm prim}}_l$, the WL contribution $C^{XY, {\rm WL}}_l$, and the detector noise $\sigma^2_{P,i}$ which is taken to be uncorrelated between different maps. The de-lensing fraction $f_{\rm DL}$ is introduced to account for the partial subtraction of the WL contribution. According to \citet{2003PhRvD..67d3001H}, the quadratic estimator method of \citet{2002ApJ...574..566H} can reduce the WL contribution to ${\tilde C}_l^{BB}$ by a factor of $7$ (implying $f_{\rm DL} =0.14$), with iterative methods promising a further reduction \citep{2003PhRvD..67d3001H}.

A scale-invariant PMF implies a scale-invariant RM spectrum \citep{2011PhRvD..84d3530P}, {\it i.~e.} the quantity 
\be
A^2_{\rm RM} = L(L+1)C_L^{\rm RM} / 2\pi
\label{a2rm}
\ee 
is constant over the scales of interest and is related to $B_{\rm SI}$ via \citep{2013PhRvD..88f3527D}
\be
A_{\rm RM}  \approx 50 \ {\rm rad/m^2} \ B_{\rm SI}/{\rm nG} \ .
\ee
The signal to noise ratio (SNR) of the detection of the primordial RM spectrum $C_L^{\rm RM,PMF}$ is given by 
\be
\left( S \over N \right)^2 = \sum_{L=1}^{L_{max}} {(f_{\rm sky}/2) (2L+1) [C_L^{\rm RM,PMF}]^2 \over [C_L^{\rm RM,PMF} + f_{\rm DG}C_L^{\rm RM,G}+\sigma^2_{{\rm RM},L}]^2} \ ,
\label{eq:ebsnrP}
\ee
where $f_{\rm DG}$ is the ``de-galaxying'' factor -- the fraction of the galactic RM spectrum that is known from other sources and can be subtracted. We use estimates of $C_L^{\rm RM,G}$ from \citet{2013PhRvD..88f3527D} based on the galactic RM map of \citet{2012A&A...542A..93O}.

\section{Detection prospects}

\begin{table*}
\centering    
\begin{tabular}{c c c c c c } 
\hline
$\nu$ (GHz) &  $\theta_{\rm fwhm}$ &  $\sigma_P$($\mu$K-arcmin) &  $(S/N)_{f_{\rm DG}=0}^{f_{\rm DL}=0}$  & $(S/N)_{f_{\rm DG}=0}^{f_{\rm DL}=1}$  & $(S/N)_{f_{\rm DG}=0.1}^{f_{\rm DL}=0}$  \\
30  &  17$'$  & 13  & 0.75 & 0.7  & 0.6  \\ 
36  &  14$'$  & 8.5 & 1    & 0.9  & 0.85 \\ 
43  &  12$'$  & 8   & 0.78 & 0.68 & 0.66 \\
51  &  10$'$  & 6.2 & 0.82 & 0.72 & 0.67 \\ 
62  &  8.2$'$ & 6   & 0.57 & 0.45 & 0.5  \\
75  &  6.8$'$ & 5.6 & 0.4  & 0.3  & 0.34 \\
90  &  5.7$'$ & 5.4 & 0.24 & 0.18 & 0.22 \\
all &  -      & -   & 2.4  & 1.95 & 1.9  \\
\hline 
\end{tabular} 
\caption{
Parameters of the 7 lowest frequency channels of PRISM \protect \citep{2013arXiv1306.2259P}, and the predicted SNR for a PMF with $B_{\rm SI}=0.1$ nG under three different assumptions about de-lensing and the galactic RM subtraction. $f_{\rm DL}=0$ means perfect removal of the WL contribution to B-modes, while $f_{\rm DL}=1$ means no WL subtraction. $f_{\rm DG}=0.1$ corresponds to removal of 90 percent of the galactic RM spectrum.
}
\label{table:prism} 
\end{table*}

Because the FR induced B-mode is proportional to the E-mode generated at last scattering, most of the RM signal comes from polarization measurements on sub-degree scales where the E-mode spectrum peaks, {\it i.~e.} $500 < l < 2000$. We find that terminating the sum in (\ref{eq:ebnoise}) at $l,l' = 2000$ includes all the informative modes. On the other hand, the mode-coupling estimator primarily probes the largest scale features of the RM -- most of the contribution to the SNR comes from $L \lesssim 30$, or angular scales of about $6^\circ$ and larger.

To demonstrate the effect of combining information from multiple channels, we consider the 7 lowest frequency channels of the recently proposed Polarized Radiation Imaging and Spectroscopy Mission (PRISM) \citep{2013arXiv1306.2259P} with parameters listed in Table~\ref{table:prism}. Adding PRISM's higher frequency channels increases the computational time with practically no improvement in the SNR. We evaluate the SNR for $B_{\rm SI}=0.1$ nG under different assumptions about the subtraction of the WL B-modes and the galactic RM. With a perfect subtraction of both ($f_{\rm DL}=f_{\rm DG}=0$), the highest detection level is 1$\sigma$ for the 36 GHz channel. However, combining the channels gives a 2.4$\sigma$ detection. This reduces to about 1.9$\sigma$ when there is no de-lensing ($f_{\rm DL}=1$) or with a 10\% residual contribution from the galactic RM ($f_{\rm DG}=0.1$). We also found that a relatively modest 25\% improvement in the resolution and the sensitivity of each of the 7 channels increases the SNR from $2.4$ to $3.9$ in the $f_{\rm DL}=f_{\rm DG}=0$ case.

It is interesting to explore how tight the constraints on PMF can become with experiments beyond PRISM. The SNR depends on the frequencies, the noise and the resolution of the channels, as well as the quality of de-lensing and galactic RM subtraction. The relative importance of the latter two depends on the experimental parameters at hand. Thus, we first discuss the dependence on $\nu_i$, $\sigma_{P,i}$ and $\theta^i_{\rm fwhm}$ for a single channel with $f_{\rm DL}=f_{\rm DG}=0$, {\it i.~e.} perfect de-lensing and de-galaxying.

The signal is determined by the FR angle, which is proportional to $B_{\rm SI}/\nu^2$. At first sight, since the signal in Eq.~(\ref{eq:ebsnrP}) is the RM {\it spectrum}, one would expect the SNR to scale as $B^2_{\rm SI}/\nu^4$. And it is certainly so for a noise-dominated measurement, {\it i.~e} when $C_L^{\rm RM,PMF} < \sigma^2_{{\rm RM},L}$ for all $L$. However, if the signal happens to be larger than the variance for some $L<L_S$, the contribution of these $L$ to the SNR in Eq.~(\ref{eq:ebsnrP}) is approximately given by 
\be 
(S/N)^2 \approx \sum_{L=1}^{L_S} f_{\rm sky} (2L+1)/2 \approx L_S^2/2 \ ,
\ee
where $L_S$ is found by setting $C_{L_S}^{\rm RM,PMF} = \sigma^2_{{\rm RM},L_S}$. For a scale-invariant RM spectrum, we use Eq.~(\ref{a2rm}) to find $L^2_S = 2 \pi A^2_{\rm RM}/ \sigma^2_{{\rm RM},L}$. For relatively small $L$ that are of interested here, $\sigma_{{\rm RM},L}$ is approximately constant and proportional to $\nu^2 \sigma_{P}$, which leads to
\be
{S \over N} \propto L_S \propto {B_{\rm SI} \over \nu^2 \sigma_P}
\label{eq:snsignal}
\ee
for the signal dominated contribution. 

To check the validity of Eq.~(\ref{eq:snsignal}) and to examine the dependence on the resolution, we take a $30$ GHz detector with $\sigma_{P,1}=1 \mu$K-arcmin and $\theta^1_{\rm fwhm}=1'$ as our reference channel, and numerically evaluate the SNR in Eq.~(\ref{eq:ebsnrP}) varying each parameter, one at a time, while keeping the other two fixed. We find that, in the signal dominated regime, the dependence of the SNR on the parameters is well approximated by
\be
{S \over N} \approx \left(28 - {\theta^i_{\rm fwhm} \over \theta^1_{\rm fwhm}} \right) \left( B_{\rm SI} \over 0.1 {\rm nG} \right) \left(30 {\rm GHz} \over \nu_i \right)^2 {\sigma_{P,1} \over \sigma_{P,i}} \ ,
\label{sn-approx}
\ee
{\it i.~e.} it agrees with Eq.~(\ref{eq:snsignal}), with the dependence on $\theta_{\rm fwhm}$ being approximately linear. Note that this expression eventually breaks down in the noise dominated regime which, in practice, means cases when $S/N<2$.

\begin{figure}
\includegraphics[height=0.5\textwidth, angle=270]{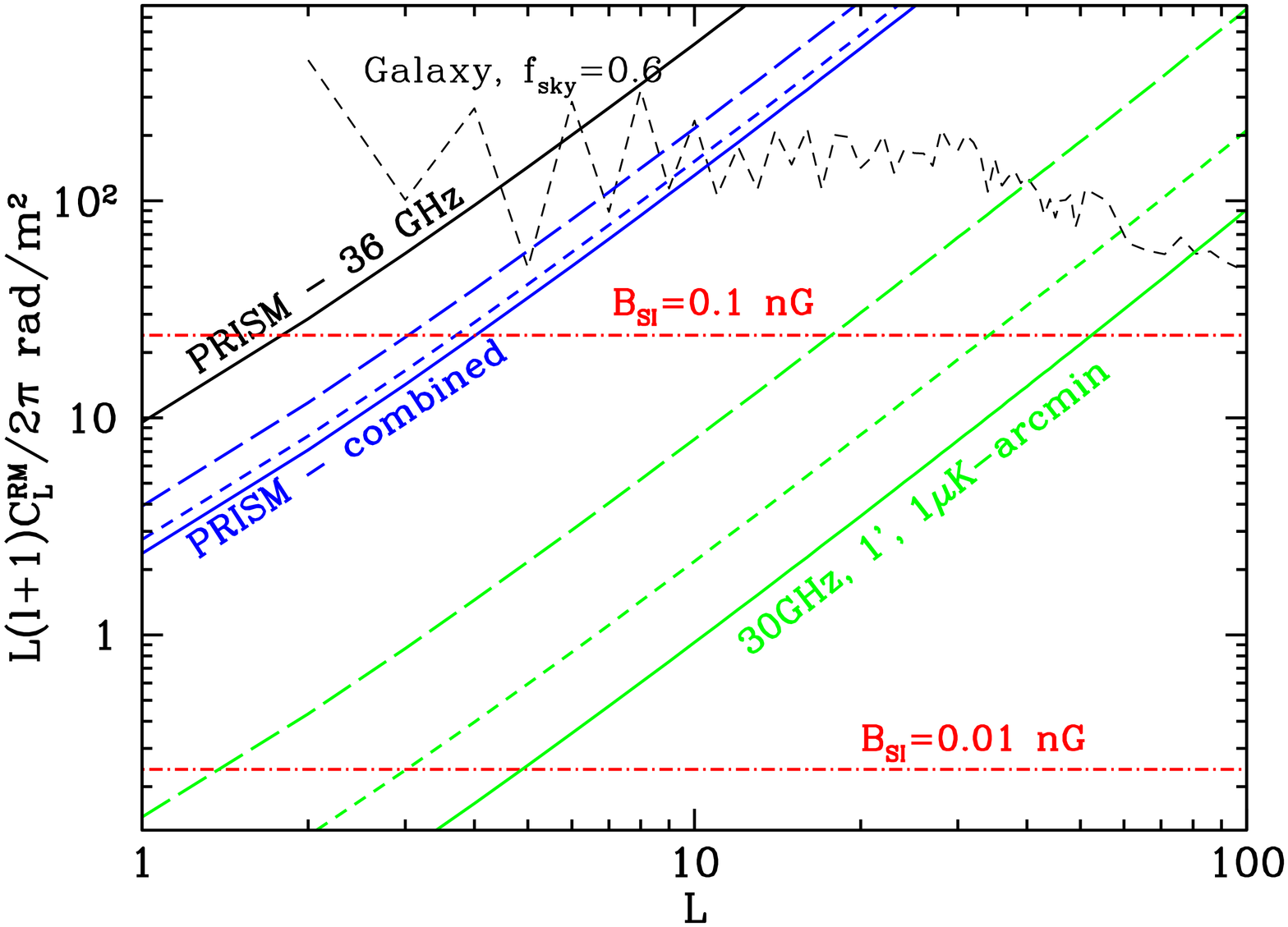}
\caption{$L(L+1)C_L^{\rm RM} / 2\pi$ for $B_{\rm SI}=0.1$ and $0.01$ nG (red dot-dash), and for the galactic RM \protect \citep{2012A&A...542A..93O,2013PhRvD..88f3527D} with $f_{\rm sky}=0.6$ (black short dash). Also shown are the variances $\sigma^2_{RM,L}$ (multiplied by $L(L+1)/ 2\pi$) of the ${\hat r}_{LM}$ estimator for the combination of PRISM's 7 lowest frequency channels (in blue) and the $30$ GHz reference channel used in Eq.~(\ref{sn-approx}) (in green) with no WL subtraction ($f_{\rm DL}=1$, long dash lines), partial WL subtraction ($f_{\rm DL}=0.14$, short dash lines) and a perfect WL subtraction ($f_{\rm DL}=0$, solid lines). To illustrate the benefit of combining channels, we also show the variance for PRISM's $36$ GHz channel, shown with a black solid line, in the case of a perfect WL subtraction.}
  \label{fig1}
\end{figure}

We show the RM spectra for $B_{\rm SI}=0.1$ and $0.01$ nG in Fig.~\ref{fig1}, along with the galactic RM spectrum based on the map of \citet{2012A&A...542A..93O} with $f_{\rm sky}=0.6$ \citep{2013PhRvD..88f3527D}. Also shown are variances $\sigma^2_{RM,L}$ (multiplied by $L(L+1)/ 2\pi$) in the estimator ${\hat r}_{LM}$ for the combination of the 7 lowest frequency channels of PRISM and the reference channel used in Eq.~(\ref{sn-approx}) under different assumptions about de-lensing. One can see that WL subtraction does not significantly affect the RM studies with PRISM, which is also apparent from numbers in Table~\ref{table:prism}. This is because of the relatively low resolution and high noise levels of PRISM's low frequency channels. However, WL subtraction becomes increasingly more important at higher resolutions and lower noise levels, as in the case of the $30$ GHz reference channel in Fig.~\ref{fig1}.

The detector noise $\sigma_P$ is inversely proportional to the square root of the length of the observation and the number of detectors used in each channel. Increasing the number of channels with comparable SNR has a similar effect, except for the benefit of having the cross-channel correlations that do not depend on the detector noise. Additionally, subtraction of various astrophysical foregrounds is greatly aided by having measurements at many frequencies. However, adding the cross-channel correlations to the same-channel correlations does not add new information about the primordial signal, since it is the same for all channels. Thus, without the foregrounds and the noise, the SNR from $N$ comparable channels would approximately scale as $\sqrt{N}$.

\section{The outlook}

Theoretically, there is no limitation on the number of channels one can employ. In principle, it is possible to use measurements of FR to lower the bounds to $B_{\rm SI} \sim 0.01$ nG or even lower. There would, of course, be serious technical challenges, not to mention the costs, when trying to fit together a significant number of high resolution low frequency channels. Also, as evident from Fig.~\ref{fig1}, it is difficult to achieve sub-nG constraints on PMF without a substantial subtraction of the galactic RM. Uncertainties in the currently available map of Oppermann et al \citep{2012A&A...542A..93O} start at 5 rad/m$^2$ near galactic poles and increase at latitudes closer to the galactic plane. Uncertainties of that order may allow for a detection of $B_{\rm SI}=0.1$ nG, but certainly not a weaker PMF. This will improve with a larger number of observed extragalactic polarized sources.

In forecasts like ours, it is usually assumed that astrophysical foregrounds will be modelled well enough to make their contribution smaller than the detector noise. However, achieving this is the main challenge in CMB polarization studies, with the frequency dependence of the known types of foregrounds used to subtract them. The relevant foregrounds are of two types \citep{2013arXiv1303.5072P}: emission from the Milky Way, which obscures CMB on large angular scales, and contamination from extragalactic sources, which affects the small scale measurements. As noted earlier, while the mode-coupling estimator probes the RM on large angular scales, the dominant contribution to the SNR comes from EB correlations on angular scales of about $11'$ ($l \sim 1000$), close to the peak of the E-mode polarization spectrum. On these scales, both the galactic emission and the extragalactic contribution are not at their strongest. One may wonder if the FR signal can be extracted by treating it as one of the frequency dependent components in the component separation algorithm. There is a reason why this may be difficult to implement. Namely, unlike other foregrounds, the FR does not generate polarization -- it rotates the polarization from existing sources. This would couple it to other polarized components in the component separation algorithm, making the relation between the components non-linear. 

One could ask if it is possible to extract the FR angle directly, by comparing polarization maps at low and high frequencies, as opposed to the mode-coupling approach discussed in this work. To the best of our knowledge, this has not been seriously investigated and we leave this for a future study.

\section*{Acknowledgments}

I thank Soma De, Bess Ng, Tanmay Vachaspati and Amit Yadav for previous collaborations and discussions that motivated this work. I acknowledge helpful conversations with Ivan Agullo, Camille Bonvin, Andrei Frolov, Andrew Jaffe, Yin-Zhe Ma and Vlad Stolyarov. My research is supported by an NSERC Discovery grant.

\label{lastpage}

\begin{thebibliography}{99}


\bibitem[\protect\citeauthoryear{Agullo 
\& Navarro-Salas}{2013}]{2013arXiv1309.3435A} Agullo, I., Navarro-Salas, J., 2013, arXiv, arXiv:1309.3435

\bibitem[\protect\citeauthoryear{Atwood et al.}{2009}]{2009ApJ...697.1071A} 
Atwood W.~B., et al., 2009, ApJ, 697, 1071, [arXiv:0902.1089 [astro-ph]] 

\bibitem[\protect\citeauthoryear{Bonvin 
\& Caprini}{2010}]{2010JCAP...05..022B} Bonvin C., Caprini C., 2010, JCAP, 5, 22, [arXiv:1004.1405 [astro-ph]] 

\bibitem[\protect\citeauthoryear{Bonvin, Caprini, 
\& Durrer}{2012}]{2012PhRvD..86b3519B} Bonvin C., Caprini C., Durrer R., 2012, PhRvD, 86, 023519, [arXiv:1308.3348 [astro-ph]]  

\bibitem[\protect\citeauthoryear{Bonvin, Caprini, 
\& Durrer}{2013}]{2013PhRvD..88h3515B} Bonvin C., Caprini C., Durrer R., 2013, PhRvD, 88, 083515, [arXiv:1112.3901 [astro-ph]] 

\bibitem[\protect\citeauthoryear{Campanelli}{2013}]{2013PhRvL.111f1301C} 
Campanelli L., 2013, PhRvL, 111, 061301 [arXiv:1304.6534 [astro-ph]] 

\bibitem[\protect\citeauthoryear{Crittenden et 
al.}{1993}]{1993PhRvL..71..324C} Crittenden R., Bond J.~R., Davis R.~L., 
Efstathiou G., Steinhardt P.~J., 1993, PhRvL, 71, 324 [arXiv:astro-ph/9303014]

\bibitem[\protect\citeauthoryear{Croft et al.}{2002}]{2002ApJ...581...20C} 
Croft R.~A.~C., Weinberg D.~H., Bolte M., Burles S., Hernquist L., Katz N., 
Kirkman D., Tytler D., 2002, ApJ, 581, 20 [arXiv:astro-ph/0012324] 

\bibitem[\protect\citeauthoryear{De, Pogosian, 
\& Vachaspati}{2013}]{2013PhRvD..88f3527D} De S., Pogosian L., Vachaspati T., 2013, PhRvD, 88, 063527 [arXiv:1305.7225 [astro-ph]] 

\bibitem[\protect\citeauthoryear{Durrer 
\& Caprini}{2003}]{2003JCAP...11..010D} Durrer R., Caprini C., 2003, JCAP, 11, 10 [arXiv:astro-ph/0305059]

\bibitem[\protect\citeauthoryear{Durrer 
\& Neronov}{2013}]{2013A&ARv..21...62D} Durrer R., Neronov A., 2013, A\&ARv, 21, 62 [arXiv:1303.7121 [astro-ph]]

\bibitem[\protect\citeauthoryear{Gluscevic et 
al.}{2012}]{2012PhRvD..86j3529G} Gluscevic V., Hanson D., Kamionkowski M., 
Hirata C.~M., 2012, PhRvD, 86, 103529 [arXiv:1206.5546 [astro-ph]]

\bibitem[\protect\citeauthoryear{Gluscevic, Kamionkowski, 
\& Cooray}{2009}]{2009PhRvD..80b3510G} Gluscevic V., Kamionkowski M., Cooray A., 2009, PhRvD, 80, 023510 [arXiv:0905.1687 [astro-ph]]

\bibitem[\protect\citeauthoryear{Grasso 
\& Rubinstein}{2001}]{2001PhR...348..163G} Grasso D., Rubinstein H.~R., 2001, PhR, 348, 163 [arXiv:astro-ph/0009061]

\bibitem[\protect\citeauthoryear{Harari, Hayward, 
\& Zaldarriaga}{1997}]{1997PhRvD..55.1841H} Harari D.~D., Hayward J.~D., Zaldarriaga M., 1997, PhRvD, 55, 1841 [arXiv:astro-ph/9608098] 

\bibitem[\protect\citeauthoryear{Hirata 
\& Seljak}{2003}]{2003PhRvD..67d3001H} Hirata C.~M., Seljak U., 2003, PhRvD, 67, 043001 [arXiv:astro-ph/0209489]

\bibitem[\protect\citeauthoryear{Hu 
\& Okamoto}{2002}]{2002ApJ...574..566H} Hu W., Okamoto T., 2002, ApJ, 574, 566 [arXiv:astro-ph/0111606]

\bibitem[\protect\citeauthoryear{Jedamzik 
\& Abel}{2011}]{2011arXiv1108.2517J} Jedamzik K., Abel T., 2011, arXiv, arXiv:1108.2517 

\bibitem[\protect\citeauthoryear{Jedamzik, Katalini{\'c}, 
\& Olinto}{2000}]{2000PhRvL..85..700J} Jedamzik K., Katalini{\'c} V., Olinto A.~V., 2000, PhRvL, 85, 700 [arXiv:1108.2517 [astro-ph]]

\bibitem[\protect\citeauthoryear{Jedamzik 
\& Sigl}{2011}]{2011PhRvD..83j3005J} Jedamzik K., Sigl G., 2011, PhRvD, 83, 103005 [arXiv:astro-ph/9911100]

\bibitem[\protect\citeauthoryear{Kahniashvili, Maravin, 
\& Kosowsky}{2009}]{2009PhRvD..80b3009K} Kahniashvili T., Maravin Y., Kosowsky A., 2009, PhRvD, 80, 023009 [arXiv:0806.1876 [astro-ph]]

\bibitem[\protect\citeauthoryear{Kahniashvili et 
al.}{2013}]{2013ApJ...770...47K} Kahniashvili T., Maravin Y., Natarajan A., 
Battaglia N., Tevzadze A.~G., 2013, ApJ, 770, 47 [arXiv:1211.2769 [astro-ph]]

\bibitem[\protect\citeauthoryear{Kamionkowski}{2009}]{2009PhRvL.102k1302K} 
Kamionkowski M., 2009, PhRvL, 102, 111302 [arXiv:0810.1286 [astro-ph]]

\bibitem[\protect\citeauthoryear{Kamionkowski, Kosowsky, 
\& Stebbins}{1997}]{1997PhRvL..78.2058K} Kamionkowski M., Kosowsky A., Stebbins A., 1997, PhRvL, 78, 2058 [arXiv:astro-ph/9609132]

\bibitem[\protect\citeauthoryear{Kosowsky 
\& Loeb}{1996}]{1996ApJ...469....1K} Kosowsky A., Loeb A., 1996, ApJ, 469, 1 [arXiv:astro-ph/9601055]

\bibitem[\protect\citeauthoryear{Kovac et al.}{2002}]{2002Natur.420..772K} 
Kovac J.~M., Leitch E.~M., Pryke C., Carlstrom J.~E., Halverson N.~W., 
Holzapfel W.~L., 2002, Natur, 420, 772 [arXiv:astro-ph/0209478]

\bibitem[\protect\citeauthoryear{Kunze 
\& Komatsu}{2013}]{2013arXiv1309.7994K} Kunze K.~E., Komatsu E., 2013, arXiv, arXiv:1309.7994 

\bibitem[\protect\citeauthoryear{Lewis}{2004}]{2004PhRvD..70d3011L} Lewis 
A., 2004, PhRvD, 70, 043011 [arXiv:astro-ph/0406096]

\bibitem[\protect\citeauthoryear{Neronov 
\& Vovk}{2010}]{2010Sci...328...73N} Neronov A., Vovk I., 2010, Sci, 328, 73 [arXiv:1006.3504 [astro-ph]]

\bibitem[\protect\citeauthoryear{Oppermann et 
al.}{2012}]{2012A&A...542A..93O} Oppermann N., et al., 2012, A\&A, 542, A93 [arXiv:1111.6186 [astro-ph]]

\bibitem[\protect\citeauthoryear{Paoletti, Finelli, 
\& Paci}{2009}]{2009MNRAS.396..523P} Paoletti D., Finelli F., Paci F., 2009, MNRAS, 396, 523 [arXiv:0811.0230 [astro-ph]]

\bibitem[\protect\citeauthoryear{Paoletti 
\& Finelli}{2013}]{2013PhLB..726...45P} Paoletti D., Finelli F., 2013, PhLB, 726, 45 [arXiv:1208.2625 [astro-ph]]

\bibitem[\protect\citeauthoryear{Planck Collaboration et 
al.}{2013a}]{2013arXiv1303.5076P} Planck Collaboration, et al., 2013, arXiv, 
arXiv:1303.5076 

\bibitem[\protect\citeauthoryear{Planck Collaboration et 
al.}{2013b}]{2013arXiv1303.5072P} Planck Collaboration, et al., 2013, arXiv, 
arXiv:1303.5072

\bibitem[\protect\citeauthoryear{Pogosian et 
al.}{2011}]{2011PhRvD..84d3530P} Pogosian L., Yadav A.~P.~S., Ng Y.-F., 
Vachaspati T., 2011, PhRvD, 84, 043530 [arXiv:1106.1438 [astro-ph]]

\bibitem[\protect\citeauthoryear{PRISM Collaboration et al.}{2013}]{2013arXiv1306.2259P} PRISM 
Collaboration, Andre, P., Baccigalupi, C., et al., 2013, arXiv, 
arXiv:1306.2259

\bibitem[\protect\citeauthoryear{Pullen 
\& Kamionkowski}{2007}]{2007PhRvD..76j3529P} Pullen A.~R., Kamionkowski M., 2007, PhRvD, 76, 103529 [arXiv:0709.1144 [astro-ph]]

\bibitem[\protect\citeauthoryear{Ratra}{1992}]{1992ApJ...391L...1R} Ratra 
B., 1992, ApJ, 391, L1 

\bibitem[\protect\citeauthoryear{Seljak 
\& Hirata}{2004}]{2004PhRvD..69d3005S} Seljak U., Hirata C.~M., 2004, PhRvD, 69, 043005 [arXiv:astro-ph/0310163]

\bibitem[\protect\citeauthoryear{Seljak, Pen, 
\& Turok}{1997}]{1997PhRvL..79.1615S} Seljak U., Pen U.-L., Turok N., 1997, PhRvL, 79, 1615 [arXiv:astro-ph/9704231]

\bibitem[\protect\citeauthoryear{Seljak 
\& Zaldarriaga}{1997}]{1997PhRvL..78.2054S} Seljak U., Zaldarriaga M., 1997, PhRvL, 78, 2054 [arXiv:astro-ph/9609169]

\bibitem[\protect\citeauthoryear{Seshadri 
\& Subramanian}{2001}]{2001PhRvL..87j1301S} Seshadri T.~R., Subramanian K., 2001, PhRvL, 87, 101301 [arXiv:astro-ph/0012056]

\bibitem[\protect\citeauthoryear{Shaw 
\& Lewis}{2010}]{2010PhRvD..81d3517S} Shaw J.~R., Lewis A., 2010, PhRvD, 81, 043517 [arXiv:0911.2714 [astro-ph]]

\bibitem[\protect\citeauthoryear{Tashiro et 
al.}{2013}]{2013arXiv1310.4826T} Tashiro H., Chen W., Ferrer F., Vachaspati 
T., 2013, arXiv, arXiv:1310.4826 

\bibitem[\protect\citeauthoryear{Turner 
\& Widrow}{1988}]{1988PhRvD..37.2743T} Turner M.~S., Widrow L.~M., 1988, PhRvD, 37, 2743 

\bibitem[\protect\citeauthoryear{Vachaspati}{1991}]{1991PhLB..265..258V} 
Vachaspati T., 1991, PhLB, 265, 258 

\bibitem[\protect\citeauthoryear{Widrow}{2002}]{2002RvMP...74..775W} Widrow 
L.~M., 2002, RvMP, 74, 775 [arXiv:astro-ph/0207240]

\bibitem[\protect\citeauthoryear{Yadav, Pogosian, \& Vachaspati}{2012}]{2012PhRvD..86l3009Y} Yadav A., Pogosian L., Vachaspati T., 2012, PhRvD, 86, 123009 [arXiv:1207.3356 [astro-ph]]

\bibitem[\protect\citeauthoryear{Yadav et al.}{2009}]{2009PhRvD..79l3009Y} 
Yadav A.~P.~S., Biswas R., Su M., Zaldarriaga M., 2009, PhRvD, 79, 123009 [arXiv:0902.4466 [astro-ph]]

\bibitem[\protect\citeauthoryear{Yadav, Shimon, \& Keating}{2012}]{2012PhRvD..86h3002Y} Yadav A.~P.~S., Shimon M., Keating B.~G., 2012, PhRvD, 86, 083002 [arXiv:1207.6640 [astro-ph]]

\end{thebibliography}
\end{document}